\documentclass[prc,preprint,showpacs,showkeys,superscriptaddress,nofootinbib,tightenlines]{revtex4}

\usepackage{graphicx,color}  %
\usepackage{bm}  %

\newcommand{\beq}{\begin{equation}}
\newcommand{\eeq}{\end{equation}}
\newcommand{\bea}{\begin{eqnarray}}
\newcommand{\eea}{\end{eqnarray}}
\newcommand{\ben}{\begin{eqnarray*}}
\newcommand{\een}{\end{eqnarray*}}

\newcommand{\simge}{\hspace*{0.2em}\raisebox{0.5ex}{$>$}
     \hspace{-0.8em}\raisebox{-0.3em}{$\sim$}\hspace*{0.2em}}
\newcommand{\simle}{\hspace*{0.2em}\raisebox{0.5ex}{$<$}
     \hspace{-0.8em}\raisebox{-0.3em}{$\sim$}\hspace*{0.2em}}

\begin{document}

\title{Three and Four Harmonically Trapped Particles\\
in an Effective Field Theory Framework}

\author{J. Rotureau}
\affiliation{Department of Physics, University of Arizona, 
Tucson, AZ 85721, USA}

\author{I. Stetcu}
\affiliation{Department of Physics, University of Washington, Box 351560, 
Seattle, WA 98195-1560, USA}

\author{B.R. Barrett}
\affiliation{Department of Physics, University of Arizona, 
Tucson, AZ 85721, USA}

\author{M.C. Birse}
\affiliation{School of Physics and Astronomy, The University of Manchester,
Manchester, M13 9PL, UK}

\author{U. van Kolck}
\affiliation{Department of Physics, University of Arizona, 
Tucson, AZ 85721, USA}


\begin{abstract}
We study systems of few two-component fermions interacting via 
short-range interactions within a harmonic-oscillator trap.
The dominant interactions, which are two-body, are organized according 
to the number of derivatives and defined in a two-body truncated model
space made from a bound-state basis. 
Leading-order (LO) interactions are solved for exactly
using the formalism of the No-Core Shell Model, 
whereas corrections are treated 
as many-body perturbations. 
We show explicitly that next-to-LO and next-to-next-to-LO interactions
improve convergence as the model space increases. 
We present results at unitarity for three- and four-fermion systems,
which show excellent agreement with 
the exact solution (for the three-body problem) 
and results obtained by others methods (in the four-body case). 
We also present results for finite scattering lengths and
non-zero range of the interaction,
including (at positive scattering length)
observation of a change in 
the structure of the three-body ground state 
and extraction of the atom-dimer scattering length.
\end{abstract}

\smallskip
\maketitle  

\section{Introduction}
\label{intro}
The details of an interparticle potential of finite range $R$
are largely irrelevant in few-body systems
when the two-body scattering length $a_2\gg R$. 
This feature, which is referred to as universality, 
and its associated simplicity
make such systems fertile testing ground for methods 
designed to tackle larger ones. 
While the measured two-nucleon scattering length happens to be large compared
to the inverse of the pion mass,
use of magnetic fields to create Feshbach resonances in 
cold, trapped atomic systems
has opened up the possibility of dialing
two-atom scattering lengths to values much larger
than the typical range of the van der Waals potential.
A recent ground-breaking achievement \cite{kohlstof} is the 
ability to further confine just a few atoms in nearly isolated sites of
optical lattices formed by laser beams. 
At low temperatures, the lattice sites may be considered as 
harmonic oscillator (HO) traps.
As long as the HO length $b$ is large, $b\gg R$, 
a trapped system still displays universal properties, 
although for $b\simle a_2$ these are rather different from 
those in a large trap or free space.

At sufficiently low energies, an effective field theory (EFT) 
has been formulated 
which uses the separation between the scales $a_2$ and $R$
to build an expansion  in powers of $R/a_2$ 
\cite{aleph,3bozos,4bozos}.
It replaces the underlying interparticle potential
by a series of contact interactions with
increasing numbers of derivatives of delta functions, 
akin to the multipole expansion in classical electrodynamics.
Except for isospin, which does play a role in the relative relevance
of few-body forces, 
the version of this EFT used in nuclear physics \cite{nukeEFTrev}
is formally indistinguishable from 
the one 
describing atomic systems \cite{Braat}. 
(Nevertheless, the underlying theories for the two cases are very different.)

As a consequence, atomic systems characterized by large scattering
lengths can be studied with techniques developed in nuclear physics
and, conversely, can provide tests for few- and many-body 
methods that can be further applied, with little or no change, to the 
description of nuclear systems at low energies. 
One such method is the No-Core Shell Model (NCSM), in which
the solution to the non-relativistic Schr\"odinger equation for 
many-nucleon systems is obtained by numerical
diagonalization using a discrete single-particle basis,
typically a HO basis \cite{tradNCSM}. 
This method is characterized by its
flexibility, being able to 
reach medium-mass nuclei with no limitation to closed or nearly-closed 
shells and, at 
the same time, handle local and non-local interactions on the same footing. 
The cost of diagonalization is controlled by restriction to a
``model space'' with a
maximum number $N^{max}$ of accessible shells above
the minimum configuration. 
The NCSM is well suited to handle EFT interactions \cite{NCSM},
which are in general non-local (in the sense of involving momenta)
and always defined up to a maximum momentum $\Lambda$, the
ultraviolet (UV) cutoff. 

In Ref.~\cite{us}, we formulated an EFT for two 
particles that support $S$-wave interactions
with $a_2\gg R$ in a HO trap of length $b\gg R$.
The EFT interactions are defined within a model space
with cutoff $\Lambda=\sqrt{2N^{max}+3}/b$.
We considered explicitly the first three orders in the expansion,
up to which point only the $S$ wave is modified by the short-range potential.
In leading order (LO), there exists only one, non-derivative 
contact interaction, which captures the physics of the scattering length
and thus reproduces the results \cite{HT} 
obtained with a pseudopotential \cite{pseudo}.
Subleading orders involve derivatives of 
contact interactions treated in perturbation theory.
The next-to-leading-order (NLO) and next-to-next-to-leading-order (NNLO) 
corrections account for the physics of the
effective range $r_2$ and lead to generalized pseudopotential results 
\cite{models,DFT_short,mehen}.
Higher orders can be calculated similarly.
We showed how convergence of the theory as function of $N^{max}$
can be improved systematically,
and is in fact better than for an exact diagonalization
of subleading orders.

In this paper we discuss systems of three and four two-state fermions 
in a trap.
For these fermions, which we can think of as having spin $s=1/2$,
the approach of Ref.~\cite{us} applies and 
we again consider explicitly the EFT to NNLO.
Thanks to the Pauli principle, three- and higher-body interactions
appear only at higher orders \cite{3stooges},
so the properties of few-body systems can be predicted.
At LO, we solve the few-body problem using the NCSM formalism,
as done before \cite{trap0}.
Beyond LO, we employ many-body
perturbation theory.

A critical aspect is the use of different values 
of the model-space cutoff for systems with different 
number of particles, which leads to improved convergence. 
This is because the 
spectator particles in a many-body system can carry some of the excitation
energy, leaving less available to an interacting pair. If the two-body and 
many-body spaces are cut off at the same number of quanta, some states of 
the two-body system will simply be omitted from the many-body space without 
their effects being taken into account by renormalization. One way to 
avoid this is to use a cutoff for the space where the many-body system is
solved that is higher than that in the 
two-body subsystem where the two-body interaction was defined.
A related approach, in which the two-body cutoff is taken 
to depend on the state of the spectators, was originally considered 
within the NCSM formalism in Ref.~\cite{NCSMcuts}, but it was later abandoned 
in favor of the simpler approach with equal cutoffs.

We calculate the lowest levels of the three- and four-fermion systems
allowing for finite values of both $a_2$ and $r_2$.
At unitarity,
our three-fermion results converge to the semi-analytical values
of Ref.~\cite{werner}. 
Our improved convergence
allows to better pinpoint the scattering length 
at which the ground-state parity changes,
a phenomenon first identified in Ref.~\cite{trap0}
and subsequently confirmed \cite{dishonest}.
We also investigate atom-dimer scattering and compare
with Ref. \cite{petrov3}. 
In the case of four fermions,
we find a ground-state energy 
similar to values in the literature
\cite{chang,other4bodies,alha}, and give an example
of an excited level.
Systems with larger numbers of fermions can be dealt with at the expense of 
more computer time.

Our approach has similarities to the one followed in Ref.~\cite{alha}, where 
an effective short-range interaction is fitted to several levels
of the pseudopotential at unitarity, but diagonalized exactly.
To the extent that they are model-independent,
results from finite-range potentials \cite{models} should be equivalent 
to ours.

The paper is organized as follows. In 
Sec.~\ref{eft_inter} we first briefly recall 
how interactions are generated using EFT 
and then outline their solution in the NCSM formalism.
In the following section, Sec.~\ref{res}, 
we show results for three and four fermions. We conclude
and discuss future applications in Sec.~\ref{last}.

\section{Formalism}
\label{eft_inter}

We consider a non-relativistic system of $A$ 
fermions of spin $s=1/2$ and mass $m$ in a HO trap of frequency $\omega$.
The HO potential can be decomposed into two pieces,
one acting on the center of mass (CM) of the $A$ particles and one
on their relative coordinates.
We denote by $\vec{r}_i$ ($\vec{p}_i$) the position (momentum) of particle $i$
with respect to the origin of the HO potential.
The Hamiltonian describing the relative motion of the particles 
is given by
\begin{eqnarray}
H= H_0+\sum_{i<j}V_{ij}+ \ldots  ,
\label{hami}
\end{eqnarray}
with
\begin{eqnarray}
H_0= \frac{1}{2m A} \sum_{i<j} (\vec{p}_i -\vec{p}_j)^2 
+\frac{m \omega ^2}{2A}\sum_{i<j}(\vec{r}_i -\vec{r}_j)^2 ,
\label{hami0}
\end{eqnarray}
$V_{ij}$ being the two-fermion interaction, 
and ``$\ldots$'' denoting three- and more-body interactions.
In the following, the zero-point of the energy scale is such that the
CM motion in the trap is omitted.

The HO introduces an energy scale $\omega$, or equivalently
a length scale, the HO length
\begin{equation}
b=\sqrt{\frac{2}{m\omega}}.
\end{equation}  
We are interested in systems where $b\gg R$, the range of the force 
between the particles.
We discuss in this section the construction and solution of $V_{ij}$ 
for few-particle systems.

\subsection{Construction of the interaction using EFT}

The interactions are constructed using EFT for the 
trapped few-fermion system.
In free space the two-body 
interaction can be characterized by its scattering phase shifts. 
For small enough values of the on-shell momentum, $k\ll 1/R$,
the phase shifts can be described by the Effective Range Expansion (ERE)
\cite{ERE}.
Potentials that generate the same values for the ERE parameters cannot be 
distinguished at low energies: 
they all lead to the same wavefunction for distances beyond the range of 
the force, $r > R$. 
This universality can be made manifest 
by taking, instead of a specific finite-range potential, 
an interaction 
expanded as a Taylor series in momentum space.

For two-state fermions, we can use the results of Ref.~\cite{us},
as long as we interpret the $\sum_{i<j}V_{ij}$ in Eq. (\ref{hami})
as a sum over pairs of particles in different states.
In this case, an $S$-wave interaction is allowed and dominates.
In the CM of the two fermions, and expressed in terms of relative coordinates,
the interactions considered in this paper are
\begin{eqnarray}
V(\vec{r}\,', \vec{r})
&=&C_0 \delta(\vec{r}\,') \delta(\vec{r})
  -C_2\left\{\left[\nabla\,'^2\delta(\vec{r}\,')\right] \delta(\vec{r})
    +\delta(\vec{r}\,') \left[\nabla^2\delta(\vec{r})\right]\right\}
         \nonumber\\
&&+ C_4 \left\{\left[\nabla\,'^4\delta(\vec{r}\,')\right] \delta(\vec{r})
	     +\delta(\vec{r}\,') \left[\nabla^4\delta(\vec{r})\right]
  + 2 \left[\nabla\,'^2\delta(\vec{r}\,')\right]
        \left[\nabla^2\delta(\vec{r})\right] \right\}
+\ldots,
\label{Taylorcoord}
\end{eqnarray}
where $C_0$, $C_2$, and $C_4$ are parameters and ``\ldots'' 
denote terms of higher orders. 
Since these interactions are singular,
a regularization procedure is introduced in form of a UV cutoff $\Lambda$.
This separates the short-distance physics,
which is not included explicitly in the dynamics of this problem at low energy,
and the 
long-distance physics, which is. 
In order for observables to be renormalization-group 
invariant,
\textit{i.e.}~independent of the arbitrary cutoff, 
the parameters $C_i$ must depend on $\Lambda$.

The HO provides a natural basis on which to expand
wavefunctions: its eigenfunctions $\phi_{nl}$, 
which can be labeled by the 
quantum numbers $n$ (radial) and $l$ (orbital), 
have energies $(N_2+3/2)\omega$, with 
\begin{equation}
N_2=2n+l.
\end{equation}
The HO basis in turn 
offers a natural cutoff 
in the form of a maximum number of shells included in the basis: 
there exists a maximum relative momentum 
\begin{equation}
\Lambda=\frac{\sqrt{2}}{b} \sqrt{N_2^{max}+3/2},
\label{Lambda}
\end{equation} 
with $N_2^{max}$ the number of quanta of
the highest energy state in the basis.
The wavefunction of the interacting system is 
a superposition of HO eigenfunctions within the model space,
determined by a solution of the Schr\"odinger equation
with the potential (\ref{Taylorcoord}).

In the case we are interested in, $a_2\gg R$, $S$-wave interactions
are enhanced by powers of $a_2$ over their natural size $R$ \cite{aleph}.
The LO interaction is the $C_0$ term
in the potential (\ref{Taylorcoord}), which represents the physics
of the scattering length $a_2$. The other terms are corrections,
in particular $C_2$ at NLO and $C_4$ at NNLO, both of which contain
the effective range $r_2$. Higher $S$-wave ERE terms and interactions
in higher partial waves appear only beyond NNLO.
To NNLO, then, the energies of the interacting system in waves with
$l\ge 1$ are not changed relative to the HO energies.

For $l=0$, on the other hand, the two-body energies $E_{2;n}$ are determined
from the solution of the Schr\"odinger equation with
a superposition of HO wavefunctions up to a maximum
$N_{2}^{max}=2 n_{max}$,
where $n_{max}$ is the largest radial quantum number carried by states in 
the two-body basis.
The solution at LO is obtained by an exact diagonalization 
considering only the $C_0$ term  in the potential (\ref{Taylorcoord}).
The parameter $C_0(N_2^{max})$ is fixed by imposing that 
one known energy,  which we take as the ground-state energy, be reproduced
at any $N^{max}_2$.
At NLO, the $C_2$ term is included in first-order perturbation theory
and $C_2(N^{max}_2)$ is fitted to a second known energy, which we take as the 
first-excited-state energy.
At NNLO, $C_2$ is treated in second-order
perturbation theory while 
the $C_4$ term is included in first-order perturbation theory,
$C_4(N^{max}_2)$ being fitted to a third 
level, which we take as the 
second-excited-state energy.

The energies used as input are 
the exact two-body energies in the
limit $N_2^{max}\to \infty$. These energies can be found by solving 
a simple transcendental equation involving 
the ERE parameters \cite{HT, models, DFT_short, mehen, us}: 
\begin{equation}
\frac{\Gamma(3/4-E_{2;n}/2 \omega)}{\Gamma(1/4-E_{2;n}/2\omega)}
=\frac{b}{2a_2}-\frac{r_{2}}{2b}\frac{E_{2;n}}{\omega}+\ldots 
 \label{eq:scat_2b}
\end{equation}
In LO, we include only the $a_2$ term;
at NLO and NNLO, we include also $r_2$.
The levels not used as input depend on $N^{max}_2$,
and for $\Lambda\simge 1/R$ they are predictions of the method.
It can be shown that including more corrections to the potential improves
the convergence to the exact solutions as 
the size of
the model space increases.
More details on the construction of the interactions 
can be found in Ref.~\cite{us}.

In this way, the two-body interaction is determined at each $N^{max}_2$
from the ERE (\textit{i.e.}~scattering) parameters, 
and can be used as input into the calculation
of $A\ge 3$ systems.
In general, for the latter we also have to include few-body interactions.
For two-state fermions, however, the Pauli principle requires
three- and more-body contact interactions to involve derivatives,
so that the extra fermions are placed in states with different 
orbital quantum numbers. The extra bodies and derivatives come with
extra factors of $R$, and the corresponding interactions
are suppressed \cite{3stooges}.
To NNLO, no few-body interaction needs to be included here.

\subsection{Many-body basis and truncation}
\label{sec:jacobi}

With $V_{ij}$ so constructed, we can predict the energy
levels of larger systems. We need to solve exactly for the LO
potential, 
and then higher-order corrections are calculated in perturbation theory, 
just as in the two-particle case \cite{us}.

In a shell-model approach, energy eigenstates are obtained by direct 
diagonalization in a many-body basis constructed with HO wavefunctions. 
There are essentially two equivalent ways to construct the basis states. 
In one, the basis states are Slater determinants constructed from 
single-particle HO wavefunctions in the lab frame. 
This leads to the wavefunction of the CM of the system factorizing 
exactly from the internal coordinates, as long as one performs 
an energy truncation of the basis states such 
that all states up to a given energy $N^{max} \omega$, 
$N^{max}$ being an integer, are included.
While antisymmetry is easily built into
this approach, the dimension of the basis space grows quickly.
None-the-less, this method can be applied efficiently to systems of
more than five particles. In the second approach, one considers states that 
depend only on internal (Jacobi) coordinates and the dependence of each of
these is described by a HO wavefunction \cite{petr}. For up to five 
particles, this is more effective than a Slater-determinant basis, 
but the antisymmetrization becomes increasingly difficult as the number of
particles grows.  As long as the same energy truncation is applied to both, 
the Slater-determinant and Jacobi bases give identical results for the 
intrinsic state of any system. 
Since in the current paper we investigate only systems with three and four 
particles, we describe in some detail the internal-coordinate approach. 
The novel truncation that we employ in the few-body system becomes more 
transparent in Jacobi coordinates.

We work with Jacobi coordinates defined in terms of differences between
the CM positions of sub-clusters 
within the $A$-body system: 
\begin{eqnarray}
\vec{\xi}_1&=&\sqrt{\frac{1}{2}}\left(\vec{r}_1-\vec{r}_2\right),
\nonumber\\
\vec{\xi}_2&=&\sqrt{\frac{2}{3}}
\left[\frac{1}{2}\left(\vec{r}_1+\vec{r}_2\right)-\vec{r}_3\right],
\nonumber\\
\vdots && 
\nonumber\\
\vec{\xi}_{A-1}&=& \sqrt{\frac{A-1}{A}}
\left[\frac{1}{A-1}\left(\vec{r}_1+\vec{r}_2+\cdots+\vec{r}_{A-1}\right)
-\vec{r}_{A}\right] .
\end{eqnarray}
Using these, the HO Hamiltonian (\ref{hami0}) can be expressed as
\begin{eqnarray}
H_0&=&\sum_{\rho=1}^{A-1} \left( \frac{{\vec{p}_{\xi_{\rho}}}^{\; 2}}{2m}
+\frac{m \omega^{2}}{2}\vec{\xi}_{\rho}^{\; 2} \right),
\end{eqnarray}
where $\vec{p}_{\xi_{\rho}}$ is the momentum canonically conjugated to 
$\vec{\xi}_{\rho}$.

To illustrate the construction of the basis,
let us consider three particles of spin $s$. 
In this case, two Jacobi coordinates $\vec{\xi}_1$, $\vec{\xi}_2$ 
are necessary to describe the internal motion.
The basis states 
\begin{equation}
{\cal A}\left\{\left[\phi_{n l}(\vec{\xi}_1) \otimes 
\phi_{{\cal NL}}(\vec{\xi}_2) \right]_{L} |(s s ){\cal S}s;S\rangle\right\}
\label{eq:basis3b}
\end{equation} 
have the spatial part constructed using HO wavefunctions 
with quantum numbers  $n$, $l$ and ${\cal N}$, ${\cal L}$ respectively, 
with the angular momentum coupled to $L$, while the spin part is constructed 
by coupling three spins $s$ to total spin $S$. 
In Eq.~(\ref{eq:basis3b}), ${\cal A}$ stands for the operator that 
antisymmetrizes the three-particle wavefunction. 
If particles 1 and 2 are already described by an antisymmetric wavefunction, 
fulfilling the 
condition
\begin{equation}
(-1)^{s+l}=1,
\end{equation}
then ${\cal A}$ ensures the correct behavior of the three-body state under  
exchange of particles 1 or 2 and 3. The basis states thus constructed are 
eigenstates of the unperturbed Hamiltonian $H_0$ and the energy of each 
three-body state can be written as $(N_3+3)\omega$ where the quantum 
number $N_3$ is defined by
\begin{eqnarray}
N_3=2n+l+2{\cal{N}}+{\cal{L}} .
\label{quanta_tot}
\end{eqnarray}
Details of the construction of a fully antisymmetrized basis from 
the states (\ref{eq:basis3b}),
the calculation of two-body matrix elements of the interaction $V_{ij}$ 
in this basis,
and the generalization to more particles and three-body forces can be found in 
Ref.~\cite{petr}. 

Because the NCSM is based on a direct numerical diagonalization, the basis must
be truncated to a computationally tractable size. 
For a system of $A$ particles, the model space is truncated by 
introducing a cutoff $N^{max}_{A}$
defined as the largest number of quanta in the eigenstates of 
$H_0$ used to construct the $A$-body  basis. 
Again using the three-body system as a concrete 
example, truncating the three-body basis at $N^{max}_3$ 
means keeping only states with HO energies 
such that $N_3=2n+l+2{\cal{N}}+{\cal{L}}\leq N_3^{max}$. 

After the truncation of the many-body space, characterized by $N_A^{max}$, 
the natural question is what would be the consistent two-body interaction. 
Since the two-body interaction is defined
in the two-body system, it is characterized by $N_2^{max}$. 
In the conventional NCSM approach, it is customary to choose the truncation 
in the two-body system so that the many-body space is the minimal required to 
include completely the two-body space. For example, if we consider just 
$S$-wave interactions, $N^{max}_2=N^{max}_3$ when one describes positive-parity
states ($N^{max}_3$ has to be even), and $N^{max}_2=N^{max}_3-1$
for negative-parity solutions ($N^{max}_3$ has to be odd). 
This was also the procedure used in Refs.~\cite{NCSM,trap0}.

However, one has to consider that the renormalization of the two-body system
means that states lying above the cutoff $N_2^{max}=2n_{max}$
have been ``integrated out" rather than simply discarded. Their effects 
are thus included implicitly in the effective two-body interaction. 
When these two interacting particles are embedded in a system with a 
larger number of particles, the spectators will carry energies associated 
with the HO levels they occupy. For example, of the $(N_3+3)\omega$ total 
energy of one of the basis states (\ref{eq:basis3b}), 
$(2{\cal N}+{\cal L}+3/2)\omega$ is carried by the relative motion of the 
spectator. As a consequence, the maximum energy available to the two-body
subsystem is smaller than that allowed by the $A$-body cutoff $N^{max}_A$
and some of the states removed by the truncation will not be accounted
for by the renormalization. 
One way to correct for this is to use the interactions renormalized with a
state-dependent two-body cutoff
$N_2^{max}=N^{max}_3-( 2 {\cal N}+{\cal L})$,
as first suggested within an NCSM approach in Ref.~\cite{NCSMcuts}.
However, the resulting state-dependent interaction is difficult to handle 
in Jacobi coordinates for systems with more than three particles, and 
cannot be incorporated in a Slater-determinant basis. In order to account
for all the two-body physics beyond our cutoff without the use of such an
interaction, we simply 
decouple the cutoff of the many-body problem from that of 
the subcluster defining any interaction. Such a prescription has some
similarity to the truncation used in Ref. \cite{alha}.

Each of our calculations is characterized by two cutoff parameters: 
$N^{max}_2$ for the two-body subsystem, 
and $N^{max}_A$ for the few-body system. 
To the order we work, no three-body forces appear and so we do not 
need to consider a separate cutoff for renormalization of a three-body
subsystem, when considering larger systems. 
Since to this order
one has to include only $S$-wave interactions,
$N_2^{max}$ is even, and 
$N^{max}_A$ is even (odd) for few-body states with even (odd) $L$.
Our final results are obtained by first increasing $N^{max}_A$
at fixed $N^{max}_2$
until they converge, and then increasing $N^{max}_2$. 
For two-body states with $N_2>N_2^{max}$, we simply set the interaction 
matrix elements to zero.
As we increase $N^{max}_A$ 
(at fixed $N_2^{max}$) from 
either $N^{max}_A=N^{max}_2$ or $N^{max}_A=N^{max}_2+1$,
we observe a rapid 
dependence on $N^{max}_A$ until it is somewhat larger than $N_2^{max}$, 
the difference reflecting the typical number of quanta carried by the
spectators. For low-lying many-body states,  
further enlarging $N^{max}_A$ makes little difference because we are 
adding only two-body states where the two-body potential
is switched off. 
Having achieved results for any observable of interest
that are stable with respect to $N^{max}_A$ for each $N_2^{max}$, we 
can then take the limit of those values for large $N_2^{max}$.
Examples are given next.

\section{Few fermions: Results}
\label{res}

In this section we present explicit results for energies of systems 
made of a few two-state fermions in a harmonic trap. 
Our goal is to show convergence as 
we increase the UV cutoff, $N^{max}_2$, 
and its systematic improvement as the order in the EFT increases.

Nothing in our method is specifically tied to unitarity
($b/a_2=0$ and $r_2/b=0$)
and we can carry out
calculations 
for finite scattering length, as well as finite range, as long as $a_2\gg r_2$.
We can in fact obtain the $b/a_2$ and $r_2/b$ dependences of any state, 
improving on the $r_2/b=0$ results of Ref. \cite{trap0} because of 
both the
accelerated convergence stemming from subleading orders
and the inclusion of the finite range of the interaction 
(which in the absence of $r_2$ arises from
the truncation of the two-body space).
In this first study of subleading orders we limit ourselves to three and
four fermions, where we can more extensively confront existing results.
Systems of more particles can be attacked with larger computer resources. 

\subsection{$A=3$ system}

We consider first a system composed of three fermions,
which allows us to illustrate the basic ideas of the method.
Before considering the more realistic
cases of finite scattering length and small effective range,
we start with the unitary case,
where semi-analytical results 
exist \cite{werner}. 
The three-fermion spectrum offers an excellent testing ground for numerical 
approaches,
including the LO considered in our previous publication \cite{trap0}
and other methods to construct effective interactions \cite{alha}.
We also use three-body energies to show how 
scattering information can be extracted for trapped few-body systems.

\subsubsection{Unitary case}

\begin{figure}[t]
\begin{center}
\includegraphics[scale=0.8,angle=-90,clip=true]{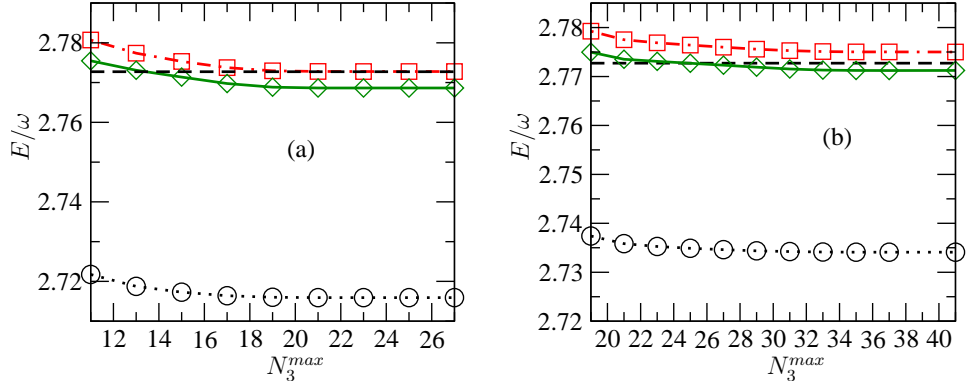}
\end{center} 
\caption{Energy in units of the HO frequency,
$E/\omega$, of the ground state $L^{\pi}=1^-$ of the $A=3$ system at 
unitarity, as function of the three-body model-space size, $N_3^{max}$:
(a) $N^{max}_2=10$; 
(b) $N^{max}_2=18$. 
(Black) Circles correspond to LO, 
(red) squares to NLO, and
(green) diamonds to NNLO. 
The (black) dashed line marks the exact value \cite{werner}.}
\label{three_fermion_evol}
\end{figure}

The ground state of the three-fermion system at $b/a_2=0$ and $r_2/b=0$
has $L=1$ and negative parity. 
Figure \ref{three_fermion_evol} shows the convergence of the energy of 
this state with the size of the 
three-body model space, $N_3^{max}$, for two values of the 
UV cutoff 
in the two-body system, (a) $N_2^{max}=10$ and (b) $N_2^{max}=18$. 
For fixed $N_2^{max}$, the Hamiltonian does not change, so that one expects 
a variational behavior of the ground-state energy with increasing the 
three-body model space. Indeed, as shown in Fig. \ref{three_fermion_evol}, 
the energy decreases until convergence is reached for a 
large enough three-body model space. 
The value of $N_3^{max}$ for which the convergence is obtained depends on the 
particular value of 
$N_2^{max}$. 
Thus, for $N_2^{max}=10$
the energy of the three-body ground state (in LO and corrections) does not 
change by more than 
$10^{-4}$ once $N_3^{max}\ge 19$, while for
$N_2^{max}=18$
convergence at this level is achieved for $N_3^{max} \ge 31$. 

\begin{figure}[t]
\begin{center}
\includegraphics[scale=0.5,angle=-90,clip=true]{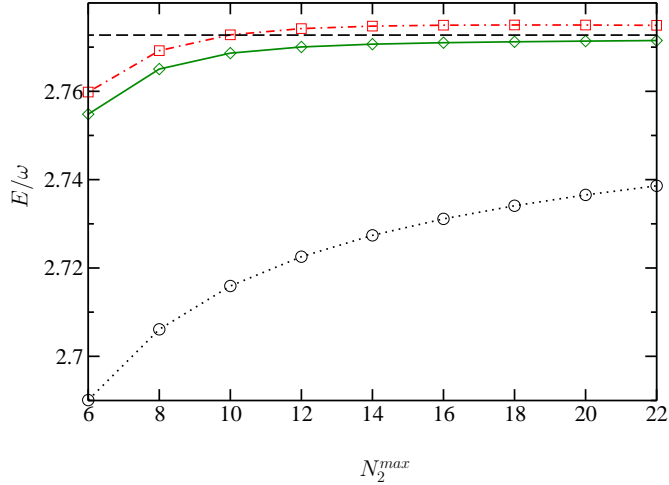}
\end{center} 
\caption{Energy in units of the HO frequency,
$E/\omega$, of the ground state $L^{\pi}=1^-$ of the $A=3$ system at 
unitarity, as function of the two-body cutoff, $N_2^{max}$.
Notation as in Fig. \ref{three_fermion_evol}.}
\label{three_fermion_per_nmax}
\end{figure}

Even though for fixed $N^{max}_2$ the errors induced by the three-body cutoff 
are eliminated, the errors induced by the truncation in the two-body sector, 
where the interaction is defined, can be eliminated either 
by taking $N^{max}_2$ to large values or by adding corrections that take into 
account physics left out by the truncation to a certain order, 
or by combination of the two.
Figure \ref{three_fermion_per_nmax} shows the convergence with respect to 
$N^{max}_2$ for 
the ground-state energy at unitarity. 
The LO calculation converges to the exact result \cite{werner},
as has been shown before in the case $N_3^{max}=N_{2}^{max}+1$ \cite{trap0}.
However, in Ref. \cite{trap0} the ground-state energy had a faster running to 
the exact value: since convergence in $N_{2}^{max}$
is from below but in $N_3^{max}$ from above, increasing $N_3^{max}$ 
at fixed $N_2^{max}$
pushes the ground-state energy further away from the exact result. 
For example, when $N_2^{max}=22$ the LO ground state is at $E/\omega=2.7413$ 
in Ref. \cite{trap0}, while the present approach gives $E/\omega=2.7386$. 
On the other hand, adding corrections to the potential speeds up the 
convergence: 
at NLO the agreement with the exact calculation is achieved faster 
than at LO, and improves still at NNLO.
For the same $N_2^{max}=22$,
the energy at NNLO is $E/\omega =2.7715$, 
very close to the exact solution 
$E/\omega=2.7727$ found in Ref. \cite{werner}. 
Thus, subleading orders provide 
significant improvement over LO results.

\begin{figure}[t]
\begin{center}
\includegraphics[scale=0.8,angle=-90,clip=true]{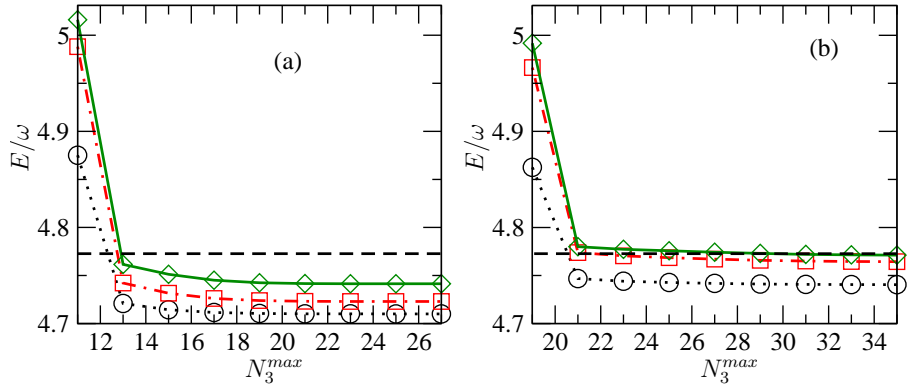}
\end{center} 
\caption{Same as Fig. \ref{three_fermion_evol}, but 
for the first excited state 
with $L^{\pi}=1^-$.}
\label{three_fermion_evol_1st_exc}
\end{figure}

\begin{figure}[t]
\begin{center}
\includegraphics[scale=0.5,angle=-90,clip=true]{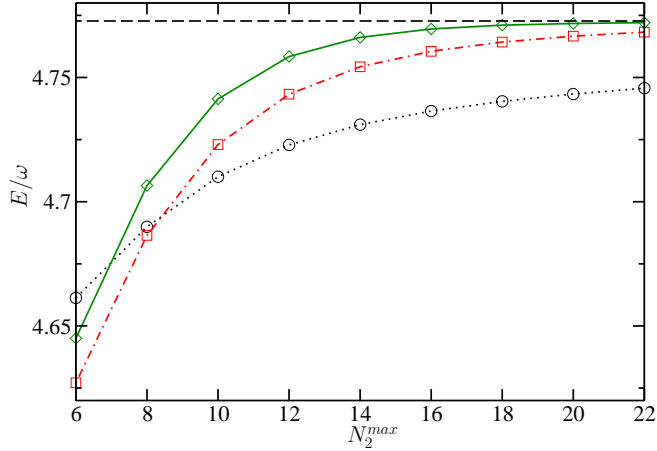}
\end{center} 
\caption{Same as Fig. \ref{three_fermion_per_nmax}, but 
for the first excited state with $L^{\pi}=1^-$.}
\label{three_fermion_per_nmax_1st_exc}
\end{figure}

The calculation of $A=3$ excited-state energies can be carried out similarly.
We show in Fig. \ref{three_fermion_evol_1st_exc} the running with the 
three-body cutoff of the energy of the first excited state 
with $L^{\pi}=1^-$,
for the same two values of $N_2^{max}$ considered before,
and in Fig. \ref{three_fermion_per_nmax_1st_exc}
the convergence with $N_2^{max}$.
The same $10^{-4}$ precision for the same two-body 
UV cutoffs considered before is achieved at somewhat larger three-body cutoffs,
$N_3^{max}\ge 23$ and $N_3^{max}\ge 35$ respectively. 
Like for the 
ground state, for a fixed $N_2^{max}$ the values 
of energies at all orders decrease until convergence is reached.
Note the sharp decrease of the energy 
as $N_3^{max}$ goes from $N_2^{max}+1$ to $N_2^{max}+3$, followed by 
small change as the three-body cutoff is further increased. 
This suggests that a small number of quanta is carried out by the spectator, 
so that most of the two-body physics can be accommodated by a relatively small 
three-body space.
The importance of having two different cutoffs in the two- and many-body 
systems clearly appears in this case. 
Indeed, if $N_{3}^{max}$ is fixed at $N_{2}^{max}+1$ one can see from 
Fig. \ref{three_fermion_evol_1st_exc} that as corrections
to the potential are added, results get worse: 
for both values of $N_{2}$
the energy at NLO and NNLO is farther away from the exact value than  
the value obtained at LO.
As Fig. \ref{three_fermion_per_nmax_1st_exc} shows,
once $N_3^{max}$ is decoupled from $N_{2}^{max}$,
the corrections to the potential again improve the energy systematically
(except at very low two-body cutoff, \textit{i.e.}~$N^{max}_2\simle 10$).
Agreement with the exact value \cite{werner} $E/\omega=4.7727$
is very good: 
for $N^{max}_2=22$,  
we find at LO $E/\omega=4.7457$, 
slightly below the value $E/\omega=4.8554$ in Ref. \cite{trap0};
at NNLO, 
$E/\omega=4.7721$. 


\begin{figure}[t]
\begin{center}
\includegraphics[scale=0.8,angle=-90,clip=true]{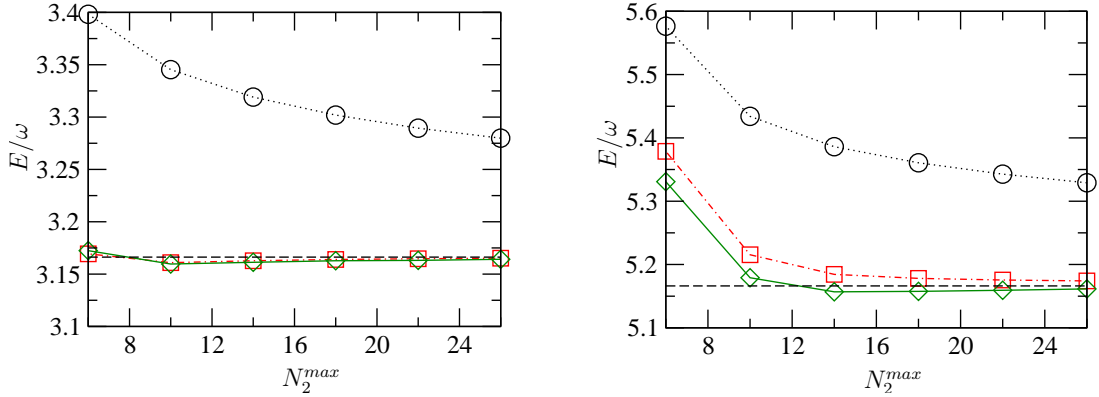}
\end{center} 
\caption{Same as Fig. \ref{three_fermion_per_nmax}, but 
for the ground state (left panel) and first excited state (right panel)
with $L^{\pi}=0^+$. }
\label{three_fermion_half_plus_per_nmax_gs_1st_exc}
\end{figure}

The general features of convergence with $N^{max}_3$ and $N_2^{max}$ 
are also shown by states in other channels,
although details vary. As an illustration, in 
Fig. \ref{three_fermion_half_plus_per_nmax_gs_1st_exc} 
we show the energies 
for the first two states with $L^{\pi}=0^+$.
In this case, the convergence with $N_2^{max}$ is not always from below.
For the lowest state, agreement with exact value \cite{werner}
$E/\omega=3.1662$ is very good
already at NLO: 
for $N^{max}_2=26$, $E/\omega=3.1652$, a difference of less than 0.05\%.
At NNLO, for the same $N_2^{max}$, the result is very close, but slightly
worse, $E/\omega=3.1641$.
For the excited state, NLO is not as good, and 
at NNLO
$E/\omega=5.1614$, within 0.1\% of the exact value \cite{werner},  
$E/\omega=5.1662$.

Overall, there is systematic improvement as $N_2^{max}$ increases, 
which is accelerated by the inclusion of higher-order interactions.

\subsubsection{Error analysis at unitarity}

Truncation of the two-body space at $N_2^{max}$ corresponds to imposing a 
two-body UV cutoff at the momentum scale (\ref{Lambda}).
The errors induced by this are expected to run as inverse powers of the cutoff 
scale, 
with a leading term of order $\Lambda^{-1}$
(see, for example, Ref. \cite{eftlore}). Since the terms in the effective 
potential represent short-range physics and thus 
must be analytic in $(p/\Lambda)^2$, only odd powers of 
$\Lambda^{-1}$ should appear. 
We therefore fit our cutoff-dependent energies with the form
\begin{equation}
\frac{E_3(N_2^{max})}{\omega}=
\frac{E_3(\infty)}{\omega}+\frac{\alpha_1}{(N_2^{max}+3/2)^{1/2}}
+\frac{\alpha_3}{(N_2^{max}+3/2)^{3/2}}+\frac{\alpha_5}{(N_2^{max}+3/2)^{5/2}}
+\ldots,
\label{eq:fitE}
\end{equation}
expressed, for convenience, in terms of dimensionless quantities. 
Here $E_3(\infty)$
is the semi-analytical result of Ref. \cite{werner} 
for the energy of the state. 
Some care is needed with this procedure since, with data for only a limited 
range of values of $N_2^{max}$, the fits can become unstable if too many terms 
are included.

The ground-state energy at LO in the range $N_2^{max}=18$ to 30 can be 
well described 
by fits with leading coefficients $\alpha_1=-0.139$ and $\alpha_3=-0.63$. 
At NNLO, 
fits to the ground state lead to values of $\alpha_1$ that are consistent 
with zero.
Setting $\alpha_1$ to zero and refitting the NNLO energy over same range 
of $N_2^{max}$ 
yields $\alpha_3=-0.30$ and $\alpha_5=-5.8$. 
The greatly improved convergence at 
NNLO that is obvious from Figs.~2 and 4 is due to elimination of the leading 
($\Lambda^{-1}$) term and reduction of the coefficient of the next one 
($\Lambda^{-3}$).

The coefficients of the higher-order terms in these expansions are large, 
implying that the series in $\Lambda^{-1}$ is rather slowly converging for the 
values of the cutoff used here. This problem appears to be worse for the 
excited 
states, where we have not been able to find stable fits 
(\textit{i.e.}~fits where the values of the coefficients do not change 
appreciably as the number of terms is increased) 
to the energies for 
$N_2^{max}\leq 30$. However one should remember that the absolute errors on 
the NNLO energies are extremely small for $N_2^{max}\geq 20$ and so this has 
no practical
consequences for our results.

\subsubsection{Finite scattering length}

\begin{figure}[t]
\begin{center}
\includegraphics[scale=0.5,angle=-90,clip=true]{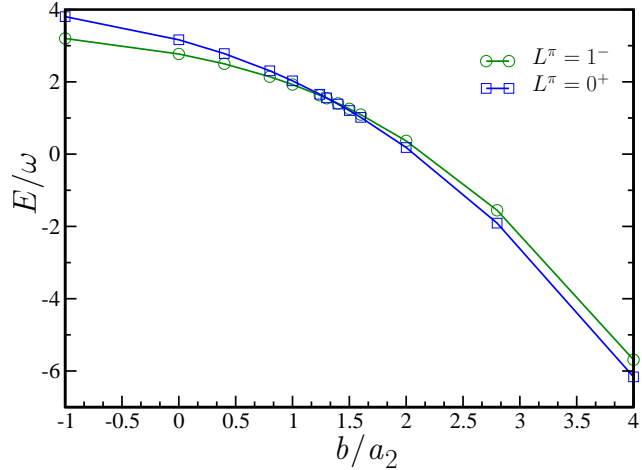}
\end{center} 
\caption{Energy in units of the HO frequency, $E/\omega$, 
of the lowest states with $L^{\pi}=1^-$ (circles) 
and $L^{\pi}=0^+$ (squares) at NNLO, as a function of the ratio $b/a_2$
between the HO length and the two-body scattering length,
when $r_2/b=0$.
At $b/a_2 \simeq 1.34$ there is an inversion of ground state.}
\label{energy_vs_b_a_plot}
\end{figure}

Away from unitarity, universality can take on a different character.
Results for the ground and first excited
state at NNLO
as a function of $b/a_2$,
but keeping $r_2/b=0$, are shown in Fig. \ref{energy_vs_b_a_plot}. 
For each value of $b/a_2$, calculations
were carried out with $N_2^{max}=22$ and the value $N_{3}^{max}$ was 
increased to reach convergence (at the same level of $10^{-4}$ as before) 
of the energy.
The values at $b/a_2=0$ are the same as in the previous
section.
As pointed out in Ref. \cite{trap0} and reproduced in
Ref. \cite{dishonest}, at positive $b/a_2$ there is an inversion
of  parity for the ground state from $L^{\pi}=1^-$, as expected from the 
HO levels with a weak interaction, to $L^{\pi}=0^+$, as expected 
from one particle moving in an $S$-wave 
around a bound state of the other two.
Here our use of higher orders in perturbation theory
allows us to pinpoint this inversion to 
$b/a_2 \simeq 1.34$, very close to our original estimate of 1.5 \cite{trap0}. 
This ground-state inversion is universal in the sense
that it holds for all potentials with negligible $r_2$.
It is one example of 
how
a trap with $b\simle a_2$ can have different universal properties than
a large, $b\gg a_2$, trap.

Positive scattering length means that the two-body ground state
is bound as the trap is removed, $b\to \infty$.
In this limit one is often interested in scattering on
the bound state.
Scattering observables are notoriously difficult to calculate in many-body 
systems, despite significant progress in this direction \cite{scat}. 
The presence of the trap, however, can simplify the calculation of low-energy 
scattering observables.
In the two-body case, given energy levels $E_{2;n}$ we can 
use Eq. (\ref{eq:scat_2b}) to extract $S$-wave scattering parameters.
In Ref. \cite{us} we examined the size of such parameters
induced by our two-body truncation, and as the present manuscript was being
written a study of the extent to which such an extraction is possible
in the two-nucleon system within the NCSM scheme appeared \cite{luu}.
Because similar connections between scattering 
properties and the spectrum of the trapped system 
exist for few-body systems,
we can use our three-body energies to extract parameters
for scattering of a particle on the two-body ground state.
A similar procedure to determine the scattering length from the full 
three-body solution was presented in Ref. \cite{PhysRevA.77.043619}.

As noted in Ref. \cite{trap0}, in the limit $b\gg  a_2$, the lowest 
three-body energy approaches the LO dimer energy, $-1/2\mu a_2^2$, 
which corresponds to the threshold for the scattering of one particle on 
the bound state of the other two. 
Indeed, if one allows the dimer to form inside a wide-enough trap,
$b \simge a_2$,
the low-lying three-body spectrum 
can be associated with the spectrum of two particles (one composite)
in a trap. 
This spectrum 
is connected to the atom-dimer low-energy parameters 
(scattering length $a_{ad}$, effective range $r_{ad}$, 
{\it etc.}) \cite{HT,models,DFT_short,mehen,us}:
\begin{equation}
 \frac{\Gamma(3/4-(E_{3;n}-E_{2;0})/2\omega)}
      {\Gamma(1/4-(E_{3;n}-E_{2;0})/2\omega)}=
\frac{b'}{2a_{ad}}-\frac{r_{ad}}{2b'}\frac{E_{3;n}-E_{2;0}}{\omega} 
+\ldots 
 \label{eq:scat_3b}
\end{equation}
Here, $E_{3;n}-E_{2;0}$ is the energy of the three-body system above 
the dimer ground state,
while $b'=1/\sqrt{\mu_{ad}\omega}$ is the HO parameter length calculated with 
the atom-dimer reduced mass, $\mu_{ad}=2m/3$. 
Note that Eq. (\ref{eq:scat_3b}) is valid only for atom-dimer relative momenta 
smaller than $1/a_2$, since $a_2$ is of the order of the size of the 
atom-dimer system, thus limiting the number of three-body states 
that can be used reliably for the extraction of atom-dimer properties. 

\begin{figure}[t]
\begin{center}
\includegraphics[scale=0.75,angle=0,clip=true]{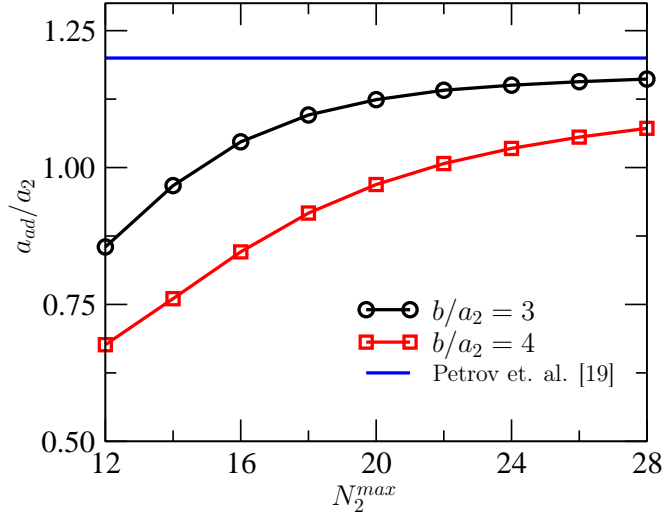}
\end{center} 
\caption{Atom-dimer scattering length 
in units of the two-body scattering length, $a_{ad}/a_2$,
as a function of the two-body cutoff, $N_2^{max}$. 
NNLO results for $b/a_2=3$ (black circles) and $b/a_2=4$ (red squares) 
are compared with the value (blue line) of Ref. \cite{petrov3}.}
\label{fig:atom_dimer_scatt}
\end{figure}

In Fig. \ref{fig:atom_dimer_scatt} we plot the atom-dimer scattering length 
$a_{ad}$ obtained at fixed $b/a_2=3$ (circles) and $b/a_2=4$ (squares). 
For each two-body cutoff, we assume that the shape 
and higher-order parameters can be neglected and, using the lowest two 
$L=0$ states 
obtained in NNLO, eliminate the atom-dimer effective range 
to obtain the scattering length. 
As shown in Fig. \ref{fig:atom_dimer_scatt}, the running is slower 
for $b/a_2=4$ than for $b/a_2=3$
because the interaction is stronger with respect to the 
HO strength. 
One thus would like smaller $b/a_2$ ratios so that better precision 
is reached for smaller cutoffs; however, for too small $b/a_2$ ratios
the dimer might not be able to form inside the trap. 
In the limit of large $N_2^{max}$ our results approach the 
continuum value of Ref. \cite{petrov3}. 

\subsubsection{Finite effective range}
As one moves away from a Feshbach resonance,
effects of the interaction range should become more pronounced and 
universality reduced
to some degree. 
Assuming a finite effective range, we can predict the changes in 
the properties of the three-body system. 
In particular, we can investigate the effect on the position of the crossing 
between the lowest $L=1$ and $L=0$ states. 

\begin{figure}[t]
\begin{center}
\includegraphics[scale=0.65,angle=0,clip=true]{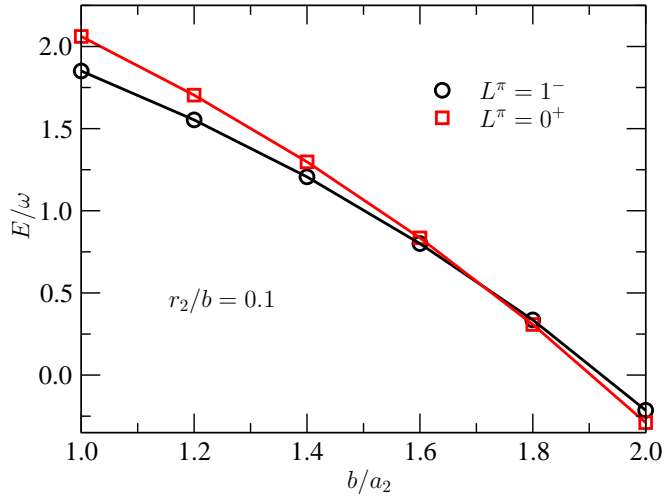}
\end{center} 
\caption{Same as Fig. \ref{energy_vs_b_a_plot}, but for $r_2/b=0.1$. 
In this case, the crossing point between the $L=1^-$ and $L=0^+$ lowest states 
moves to $b/a_2\simeq 1.75$, a larger value than in the absence of effective 
range.}
\label{energy_vs_b_a_plot_r01}
\end{figure}

In Fig. \ref{energy_vs_b_a_plot_r01} we plot the corresponding energies 
as a function of $b/a_2$, for $r_2/b=0.1$. 
As one can see, such an $r_2$ changes each of the energies 
by less than $\omega$ in the interval displayed.
One consequence is a change in the 
position where
the $L=0$ becomes lower in energy than the $L=1$ state,
which is now at $b/a_2\simeq 1.75$.
Indeed, the effect of the finite positive range is a shift of the crossing 
point to larger values of $b/a_2$.

\subsection{$A=4$ system}
We now consider the system made of four two-component fermions in a trap. 
As before, we fix the value of 
the two-body cutoff $N_2^{max}$ and increase the size of the many-body model 
space, 
defined here as $N_4^{max}$. We show only results
at unitarity, although finite scattering length and effective range
can be entertained as well. Of course, because
of the larger number of particles we limit ourselves to smaller spaces.

\begin{figure}[t]
\begin{center}
\includegraphics[scale=0.8,angle=-90,clip=true]{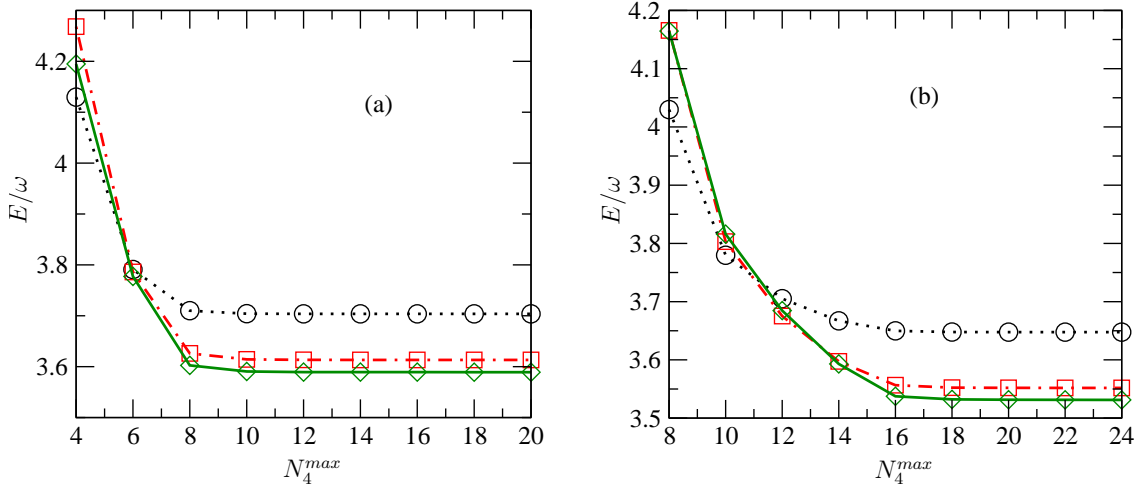}
\end{center} 
\caption{Energy in units of the HO frequency,
$E/\omega$, of the ground state $L^{\pi}=0^+$ of the $A=4$ system at 
unitarity, as function of the four-body model space size, $N_4^{max}$:
(a) $N^{max}_2=4$;
(b) $N^{max}_2=8$. 
Notation as in Fig. \ref{three_fermion_evol}.}
\label{four_fermion_evol_bis}
\end{figure}

For states with zero total angular momentum,
the smallest value for the four-body cutoff is 
$N_4^{max}=N_2^{max}$. Results for the convergence of 
the ground-state energy at unitarity with respect to $N_4^{max}$ are plotted 
in Fig. \ref{four_fermion_evol_bis} for two values of the two-body cutoff, 
(a) $N_2^{max}=4$ and (b) $N_2^{max}=8$. 
As for the $A=3$ ground state, at each order convergence is from above.
For the two values of $N_2^{max}$ displayed, we notice the same rather 
sharp decrease of the energy as for the $A=3$, $L=1^-$ first-excited state
shown in Fig. \ref{three_fermion_evol_1st_exc}.
As in the latter case, improvement with order is visible only
after this sharp decrease.
Here near convergence 
is achieved when $N_4^{max}$ reaches
$N_2^{max}+4$ for $N_2^{max}=4$ and 
$N_2^{max}+8$ for $N_2^{max}=8$,
suggesting that more energy is taken away from the two-body subsystems
than in the three-body case.

\begin{figure}[t]
\begin{center}
\includegraphics[scale=0.8,angle=-90,clip=true]{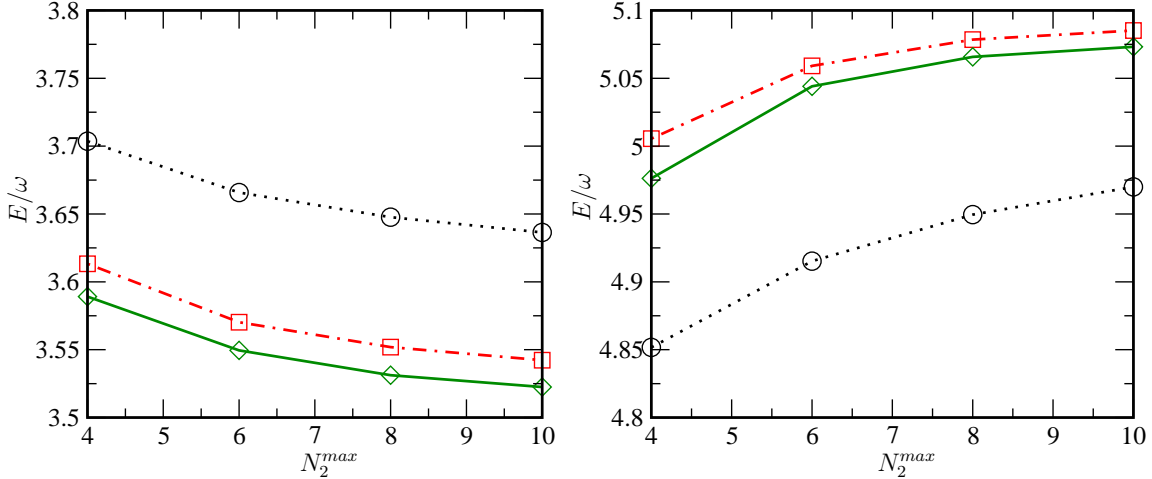}
\end{center} 
\caption{Same as Fig. \ref{three_fermion_per_nmax}, but 
for the ground state (left panel) and first excited state (right panel)
with $L^{\pi}=0^+$ for
the $A=4$ system at unitarity.
Notation as in Fig. \ref{three_fermion_evol}.}
\label{four_fermion}
\end{figure}

Figure \ref{four_fermion} shows the convergence in $N^{max}_2$ for the ground 
and first-excited states at unitarity. For 
each cutoff $N^{max}_2$, the four-body model space was increased
until convergence. 
In LO, the ground-state energy for $N^{max}_2=10$
is $E/\omega =3.64$, 
to be compared with $E/\omega =4.01$ in Ref. \cite{trap0}.
With corrections up to NNLO
$E/\omega = 3.52$, which 
is in good agreement with previous calculations where the 
ground-state energy was found to be 
$3.6 \pm 0.1$ \cite{chang},
$3.551 \pm 0.009$ \cite{other4bodies},
and $3.545 \pm 0.003$ \cite{alha}. 
Like in the case of three particles, improvement is significant 
and systematic.

\section{Conclusions and Outlook}
\label{last}
We have considered systems 
made out of a few two-component fermions in a HO trap of length $b$
using interactions generated by the
application of effective field theory up to NNLO.
To this order, the interactions are purely two-body 
and determined by the two-body scattering length $a_2$ and effective
range $r_2$. 

Calculations at LO were performed by solving
the many-body Schr\"odinger equation via a direct diagonalization, 
similar to the NCSM approach,
whereas higher-order corrections were treated as perturbations. 
We have seen that, as in the two-body case \cite{us}, convergence
is sped up by adding corrections to the potential. 
We have shown the necessity of using different values for the two-
and many-body UV cutoffs in order to allow for enough two-body states
in the many-body environment to match the two-body physics included
in the construction of the interaction.

By doing so, results at unitarity for the three- and four-fermion systems 
agree very well with 
other solutions, either semi-analytical \cite{werner}
or using other numerical methods \cite{chang,alha,other4bodies}.
We also have presented results for finite values of 
$a_2$ and $r_2$.
We were able to more precisely determine the ratio $b/a_2$
where, with vanishing effective range, there is
a parity inversion in the ground state of the three-body system \cite{trap0}.
If the range is known, its effect on the location of this point
can be calculated.
On either side of this transition the ground state is universal in the
sense of being the same regardless of the details of the potential,
as long as its range $R\ll a_2$. However, the existence of the transition
shows that the trap can lead to qualitatively different behavior
compared to the free system.

Above the transition point, the ground state is the one expected
from one atom moving in a $S$-wave around a dimer.
We were able to use our calculated three-body energy levels
to obtain an estimate for the atom-dimer scattering length
comparable to the value found in Ref. \cite{petrov3}.

This work can be extended in various directions.
First, as more data on trapped few-fermion systems \cite{kohlstof}
appear, one could determine $r_2$ for specific atoms
and predict its effects on the few-body dynamics. 
Second, one can apply the same method to other systems, such
as bosons or fermions with more components. 
In these cases a three-body force appears already at LO \cite{3bozos},
whose parameter needs to be determined from the three-body system itself.
The corresponding UV limit-cycle behavior \cite{3bozos}
is expected to survive the presence of the trap.
A system of particular interest is the atomic nucleus, where $r_2$ is known
and the LO three-nucleon force can be determined 
either from the triton binding energy \cite{NCSM}
or from
the neutron-deuteron scattering length 
(through
the lowest energy levels of the trapped system of two neutrons and one proton).
We can now calculate such systems systematically to high orders.

\section*{Acknowledgments}
We thank Calvin Johnson and Dick Furnstahl for useful discussions,
and especially 
Petr Navr\'atil for providing us with a code to solve the four-body system.
The work reported here benefited from hospitality extended to its
authors by the National Institute for Nuclear Theory 
at the University of Washington,
and to UvK by the Kernfysisch Versneller Instituut at the
Rijksuniversiteit Groningen.
This research was supported in part by 
the US NSF under grants PHY-0555396 and PHY-0854912 (BRB, JR), 
by the US DOE under grants DE-FG02-04ER41338 (JR, UvK)
and  DE-FC02-07ER41457 (IS),
and by the UK STFC under grants PP/F000448/1 and ST/F012047/1 (MCB).

\end{document}